\documentclass[a4paper]{jpconf}
\usepackage{graphicx}
\usepackage{epsfig}

\def\oo{\'o}

\def\as{\alpha_s}
\def\asz{\as(\mz)}
\def\mz{M_Z}
\def\q2{Q^2}
\def\asmz#1#2#3#4#5#6{\asz = #1\pm #2\ {\rm (stat.)}\ ^{+#4}_{-#3}\ {\rm (exp.)}\ ^{+#6}_{-#5}\ {\rm (th.)}}
\def\g2{GeV$^2$}
\def\etjb{E^{\rm jet}_{T,{\rm B}}}

\def\figdir{./}

\begin{document}
\title{Precision measurements of $\as$ at HERA\footnote{Talk given in the
    ``HEP2007 International Europhysics Conference on High Energy
    Physics'' (Manchester, UK, July 2007).}}

\author{Claudia Glasman\footnote{Ram\oo n y Cajal Fellow.} (on behalf
  of the H1 and ZEUS Collaborations)}

\address{Universidad Aut\oo noma de Madrid, Spain}

\ead{claudia@mail.desy.de}

\begin{abstract}
Recent determinations of $\asz$ from the H1 and ZEUS Collaborations
using inclusive-jet cross-section measurements in neutral current deep
inelastic scattering at high $\q2$ are presented. A combined value of 
$\asz=0.1198\pm 0.0019\ ({\rm exp.})\pm 0.0026\ ({\rm th.})$ was
obtained from these measurements. The determinations of $\as$ at
various scales clearly show the running of the coupling from HERA jet
data alone and in agreement with the prediction of QCD.
\end{abstract}

\section{Introduction}
\begin{figure}[th]
\setlength{\unitlength}{1.0cm}
\begin{picture} (18.0,6.8)
\put (0.0,-0.5){\epsfig{figure=\figdir 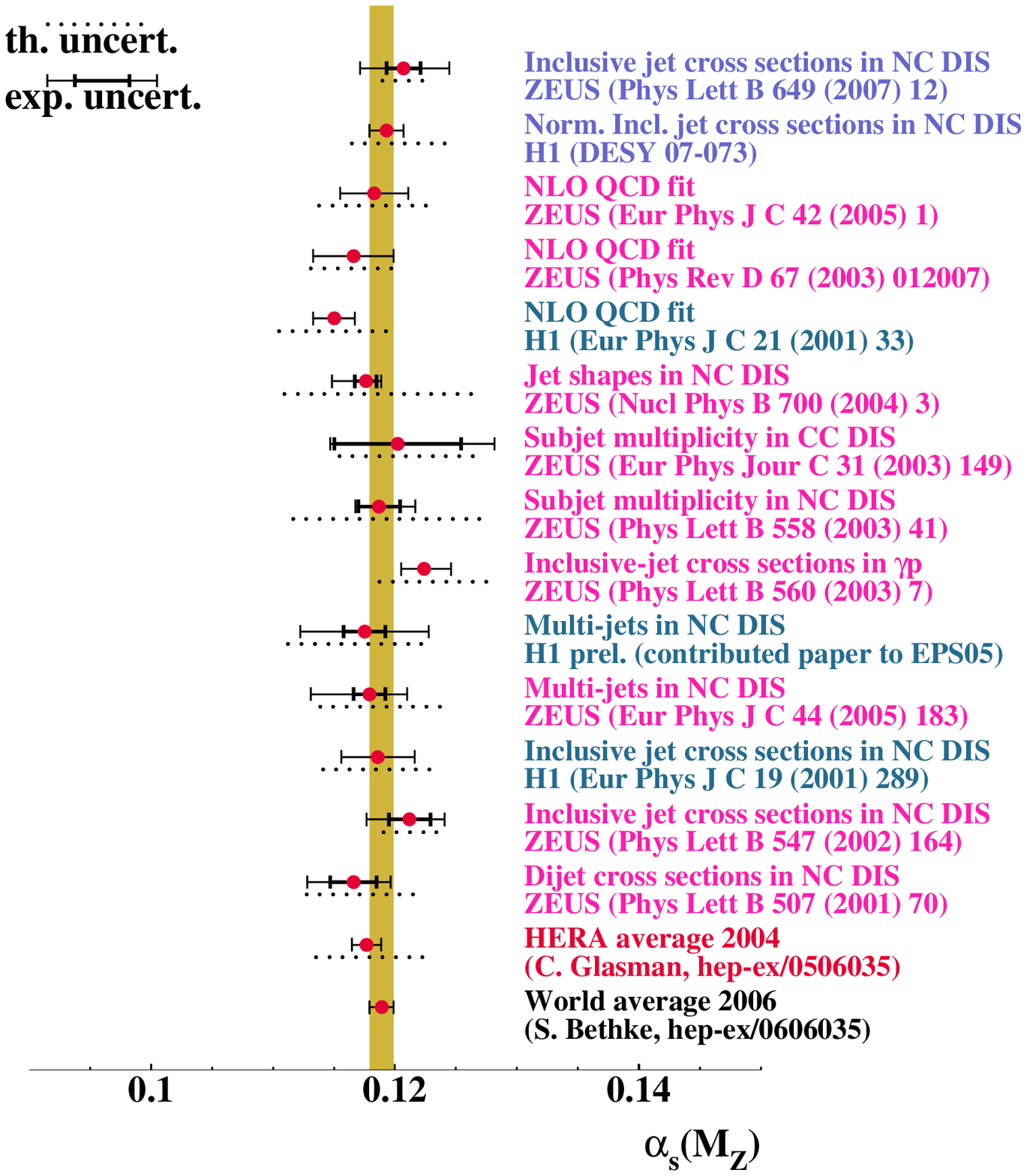,width=8cm}}
\put (7.0,-0.5){\epsfig{figure=\figdir 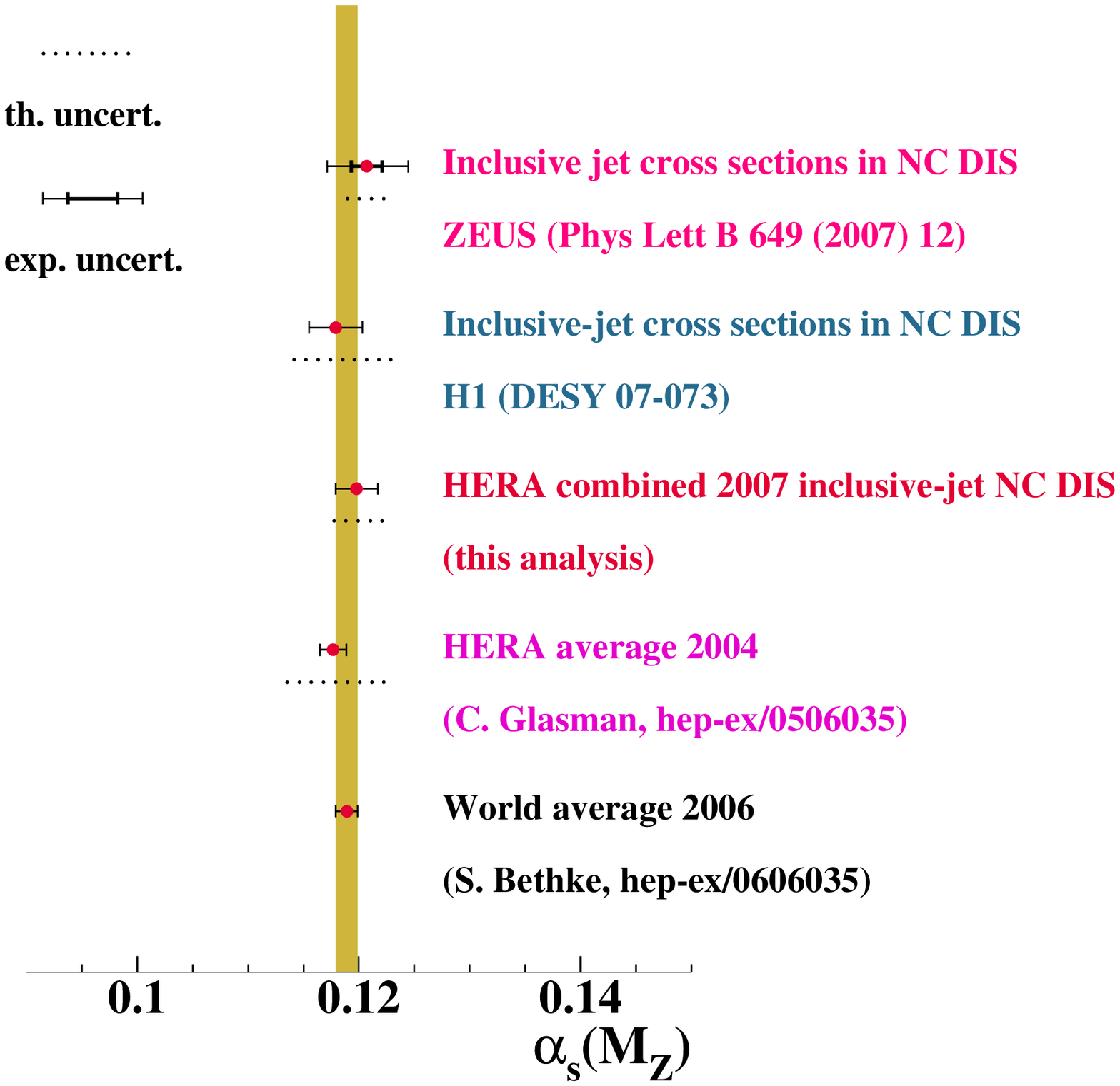,width=8cm}}
\put (2.7,-0.3){\bf\small (a)}
\put (9.8,-0.3){\bf\small (b)}
\end{picture}
\caption{\label{fig1}
{
$\asz$ determinations from the ZEUS and H1 Collaborations. The HERA
combined 2007 $\asz$ and the HERA-2004 and world-2006 averages are
also shown.
}}
\end{figure}

The strong coupling constant, $\as$, is one of the fundamental
parameters of QCD. However, its value is not predicted by the theory
and must be determined experimentally. The success of perturbative QCD
is strengthened by precise and consistent determinations of the
coupling from many diverse phenomena such as $\tau$ decays, event
shapes, $Z$ decays, etc. At HERA, many precise determinations of $\as$
have been performed from a variety of observables based on jets and on
structure functions. A summary of the determinations of $\asz$ done by
the H1 and ZEUS Collaborations is shown in Fig.~\ref{fig1}a. The
values are consistent with each other and in good agreement with the
HERA-2004~\cite{hep-ex/0506035} and world-2006~\cite{hep-ex/0606035}
averages. New determinations of $\asz$, recently published by the
ZEUS~\cite{inclusive} and H1~\cite{h1} Collaborations, are
presented. The new HERA-2007 $\asz$ combined value is also presented.

\section{Determinations of $\asz$ from the H1 and ZEUS Collaborations}
New determinations of $\as$ have been recently published by the
H1~\cite{h1} and ZEUS~\cite{inclusive} Collaborations. These
determinations have been performed from the measurements of
inclusive-jet cross sections in neutral current (NC) deep inelastic
scattering (DIS) at high $\q2$. The procedure to determine $\as$ from jet
observables used by ZEUS is based on the $\as$ dependence of the
calculations and takes into account the correlation with the parton
distribution functions (PDFs). The method consists of performing
next-to-leading-order (NLO) calculations using sets of PDFs for which
different values of $\asz$ were assumed in the fits. A
parameterisation of the $\as$ dependence of the theory for the given
observable is obtained and then the value of $\asz$ is extracted from
the measured cross section using such a parameterisation. This procedure
handles correctly the $\as$ dependence of the calculations and
preserves the correlation with the PDFs. A similar method is used by
the H1 Collaboration.

The new determinations of $\asz$ focus on obtaining the most precise
values. ZEUS has determined a value of $\asz$ with the aim of
decreasing the theoretical uncertainties. From the measured cross
section as a function of $\q2$ for $\q2>500$~\g2, the 
value~\cite{inclusive}

\vspace{0.3cm}
\centerline{$\asmz{0.1207}{0.0014}{0.0033}{0.0035}{0.0023}{0.0022}$}
\vspace{0.3cm}

\hspace{-0.6cm}
has been determined. The experimental uncertainty is dominated by the
jet energy scale and amounts to $2\%$. The theoretical uncertainties
comprise the terms beyond NLO, which are the dominant contribution,
and the uncertainties due to the proton PDFs and hadronisation
effects. This value of $\asz$ is very precise, with a total
uncertainty of $3.6\%$ and a contribution of only $1.9\%$ from the
theoretical uncertainty. This minimisation of the theoretical
uncertainty comes from the optimisation of the phase-space region
selected by ZEUS: the theoretical uncertainties arising from terms
beyond NLO and the PDFs decrease as $\q2$ increases.

The H1 Collaboration has extracted the value~\cite{h1}

\vspace{0.3cm}
\centerline{$\asz=0.1193\pm 0.0014\ {\rm (exp.)}_{-0.0034}^{+0.0050}\ {\rm (th.)}$}
\vspace{0.3cm}

\hspace{-0.6cm}
from the normalised double-differential inclusive-jet cross sections
in the region $150<\q2<15000$ \g2. In this way, a
region of phase space is selected for which experimental uncertainties
are well under control. Furthermore, the use of a normalised cross section
allows a partial cancellation of correlated uncertainties. The
experimental uncertainties are dominated by the jet energy scale and
the model dependence of the correction factors. This analysis also
gives a very precise value of $\asz$, with a total uncertainty of
$4.3\%$, and a contribution of only $1.2\%$ from the experimental
uncertainty. The advantage of using normalised cross sections can be
appreciated by comparing this determination with that obtained from
using double-differential inclusive-jet cross sections in the same
$\q2$ range:
$\asz=0.1179\pm 0.0024\ {\rm (exp.)}_{-0.0043}^{+0.0059}\ {\rm (th.)}$.
Another test was done by restricting $\q2$ to the range between 700
and 5000 \g2. The result, 
$\asz=0.1171\pm 0.0023\ {\rm (exp.)}_{-0.0014}^{+0.0034}\ {\rm
  (th.)}$, benefits from a significant decrease of the theoretical
uncertainties at the expense of an increase in the experimental
uncertainty. This trend is compatible with the result obtained by ZEUS.

\begin{figure}[th]
\setlength{\unitlength}{1.0cm}
\begin{picture} (18.0,6.8)
\put (0.0,-1.0){\epsfig{figure=\figdir 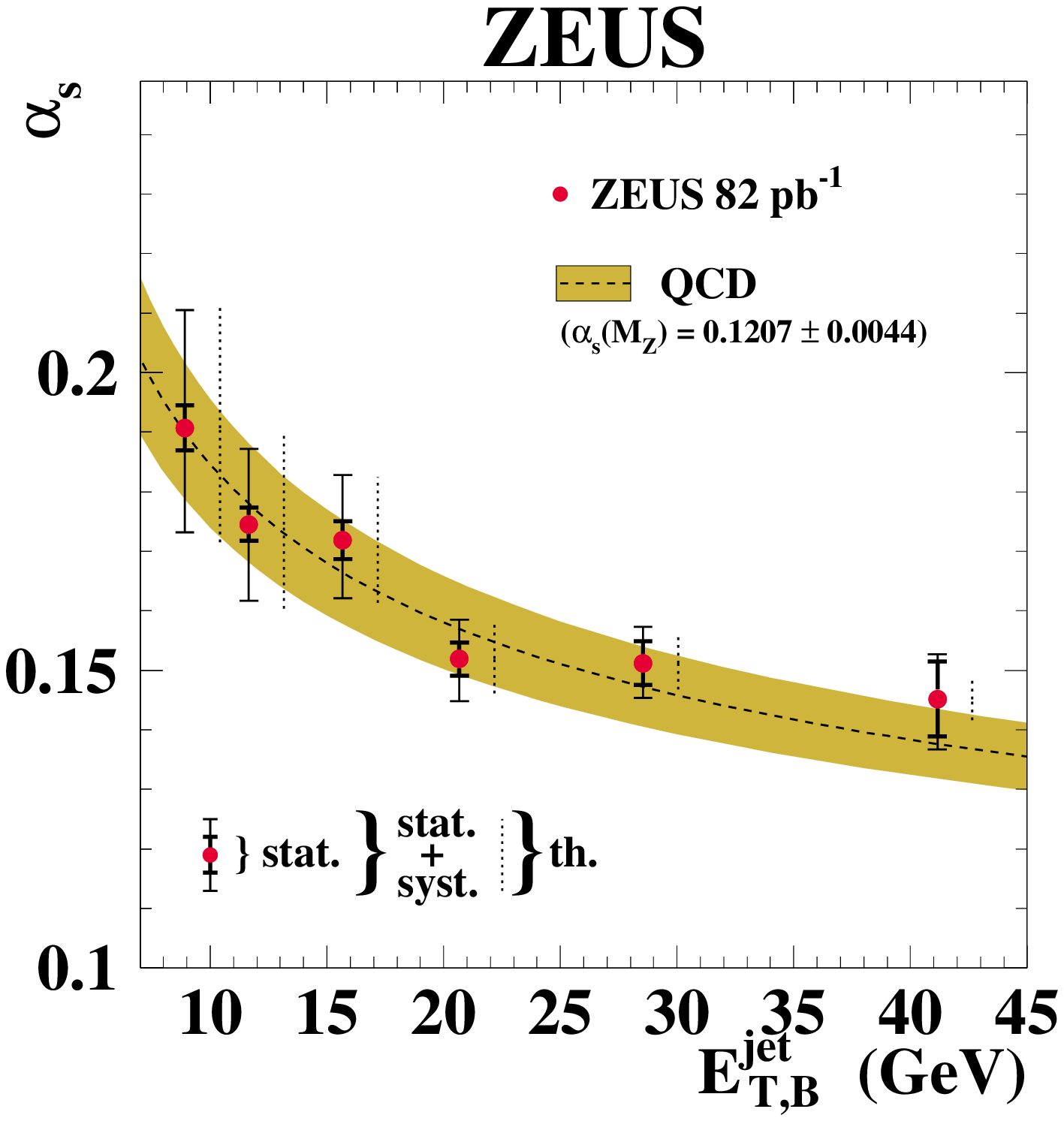,width=8cm}}
\put (7.0,-0.5){\epsfig{figure=\figdir 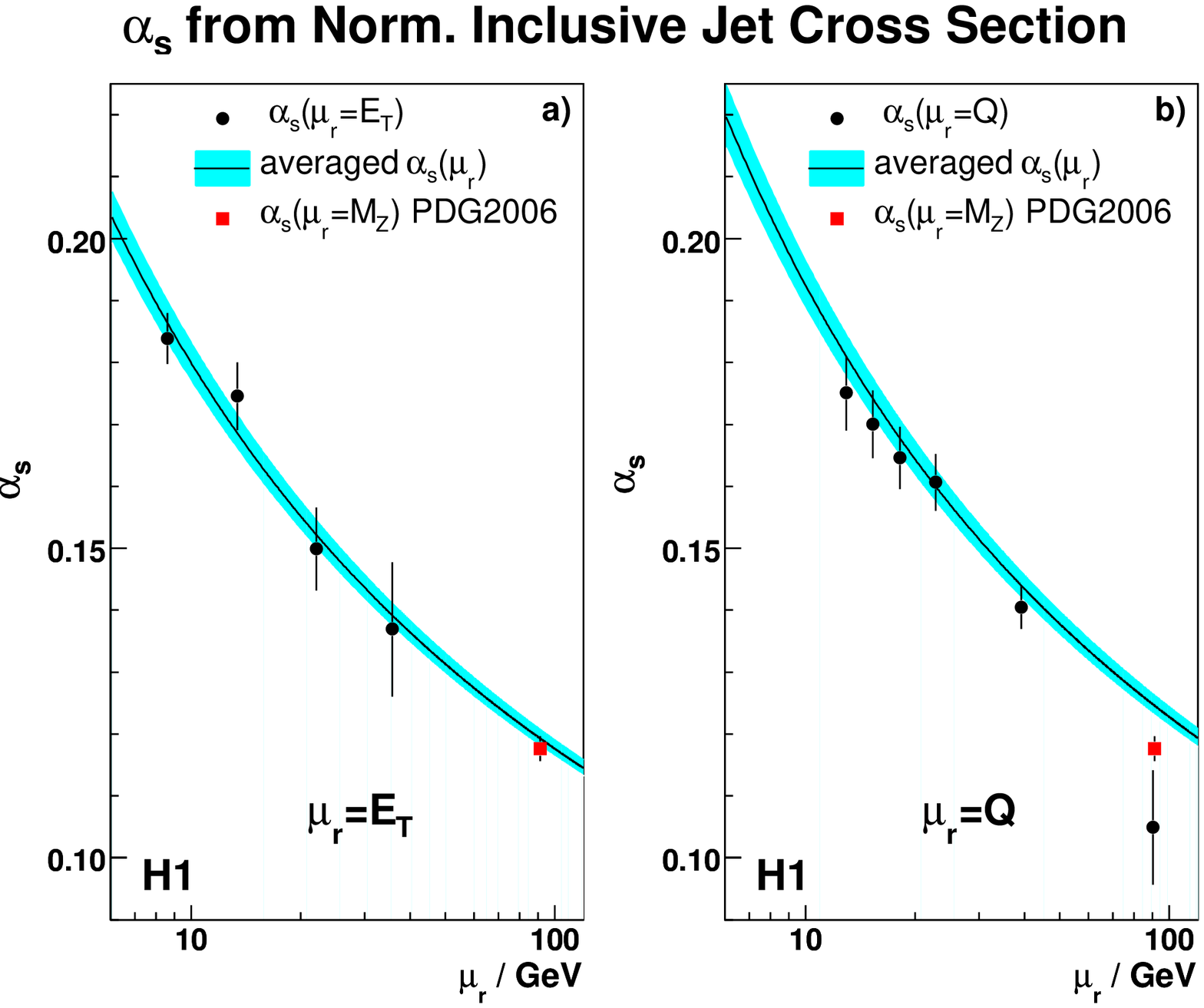,width=8cm}}
\end{picture}
\caption{\label{fig2}
{
$\as$ determinations from the ZEUS and H1 Collaborations as a function
of the scale.
}}
\end{figure}

\begin{figure}[th]
\setlength{\unitlength}{1.0cm}
\begin{picture} (18.0,5.3)
\put (0.0,-1.5){\epsfig{figure=\figdir 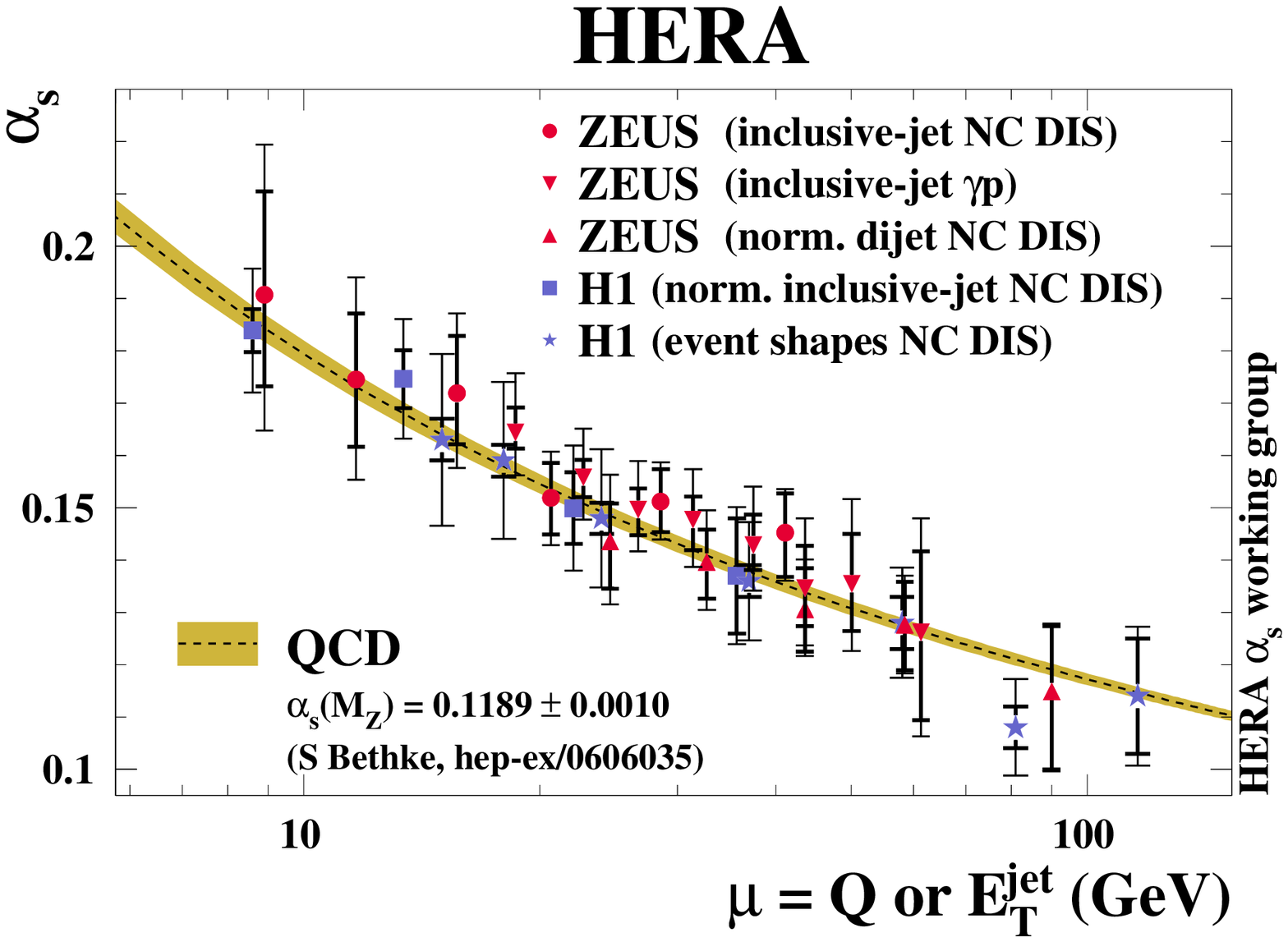,width=8cm}}
\put (8.0,-1.5){\epsfig{figure=\figdir 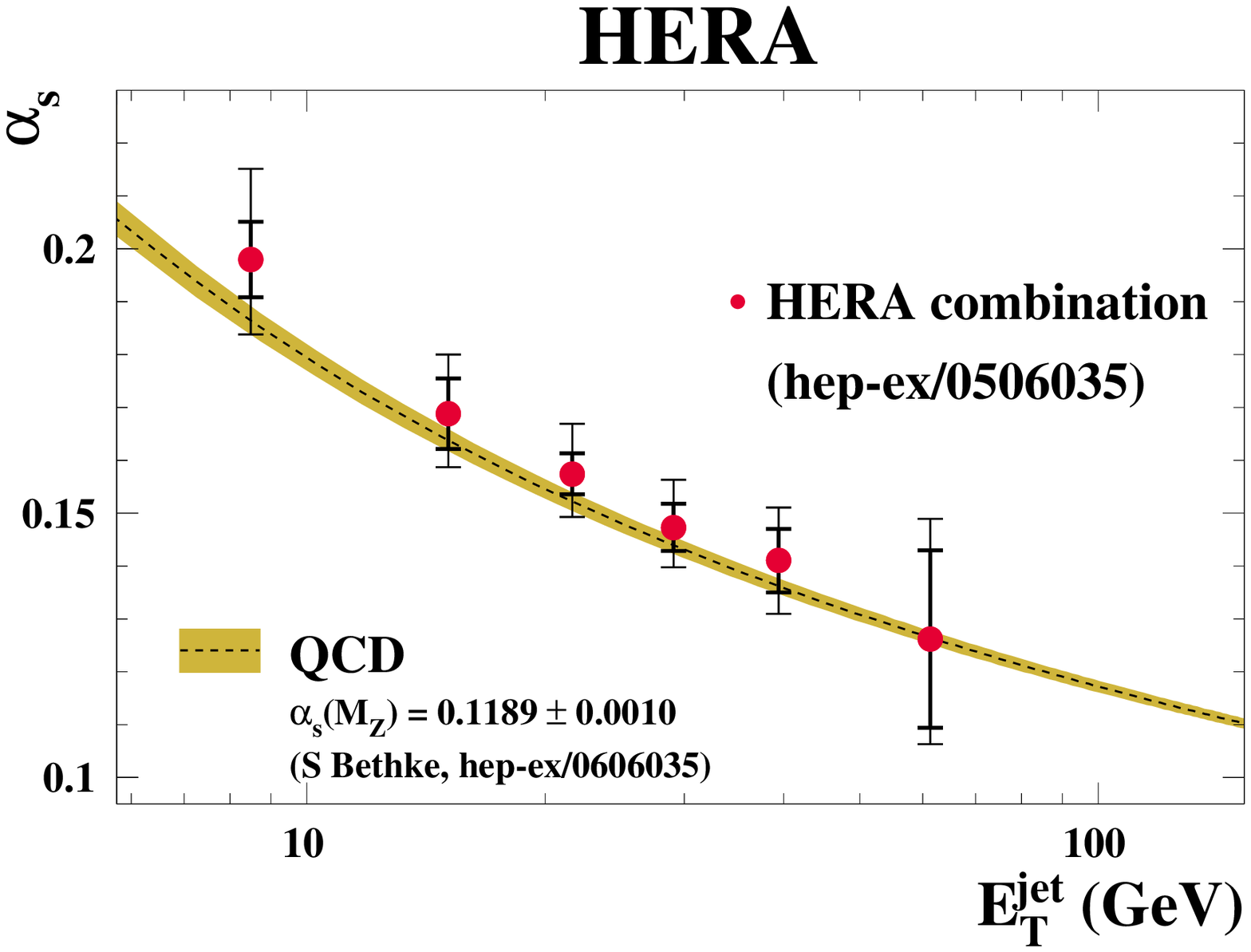,width=8cm}}
\put (4.0,0.0){\bf\small (a)}
\put (12.5,0.0){\bf\small (b)}
\end{picture}
\caption{\label{fig3}
{
(a) $\as$ determinations from the H1 and ZEUS Collaborations as a
function of the scale. (b) HERA combined values of $\as$ as a function
of $\etjb$.
}}
\end{figure}

The energy-scale dependence of $\as$ has been tested by both
Collaborations. ZEUS has determined $\as$ from the measured
inclusive-jet cross section as a function of the jet transverse energy
in the Breit frame, $\etjb$, at different values of
$\etjb$~\cite{inclusive}. H1 has tested the energy-scale dependence 
of $\as$ by determining the coupling from the normalised inclusive-jet
cross sections both as a function of $\etjb$ and $Q$. In all cases,
the results (see Fig.~\ref{fig2}) show a decrease of $\as$ as the
energy scale increases and are in good agreement with the running of
$\as$ as predicted by QCD over a large range of the
scale. Figure~\ref{fig3}a shows a compilation of all the available
measurements of $\as$ as a function of the energy scale at HERA. The
values of $\as$ at different scales determined from jet observables
have been combined~\cite{hep-ex/0506035} (see Fig.~\ref{fig3}b): the
running of $\as$ is observed from HERA jet data alone.

\section{HERA combined 2007 $\asz$}
\begin{figure}[th]
\setlength{\unitlength}{1.0cm}
\begin{picture} (18.0,7.5)
\put (0.0,-0.3){\epsfig{figure=\figdir 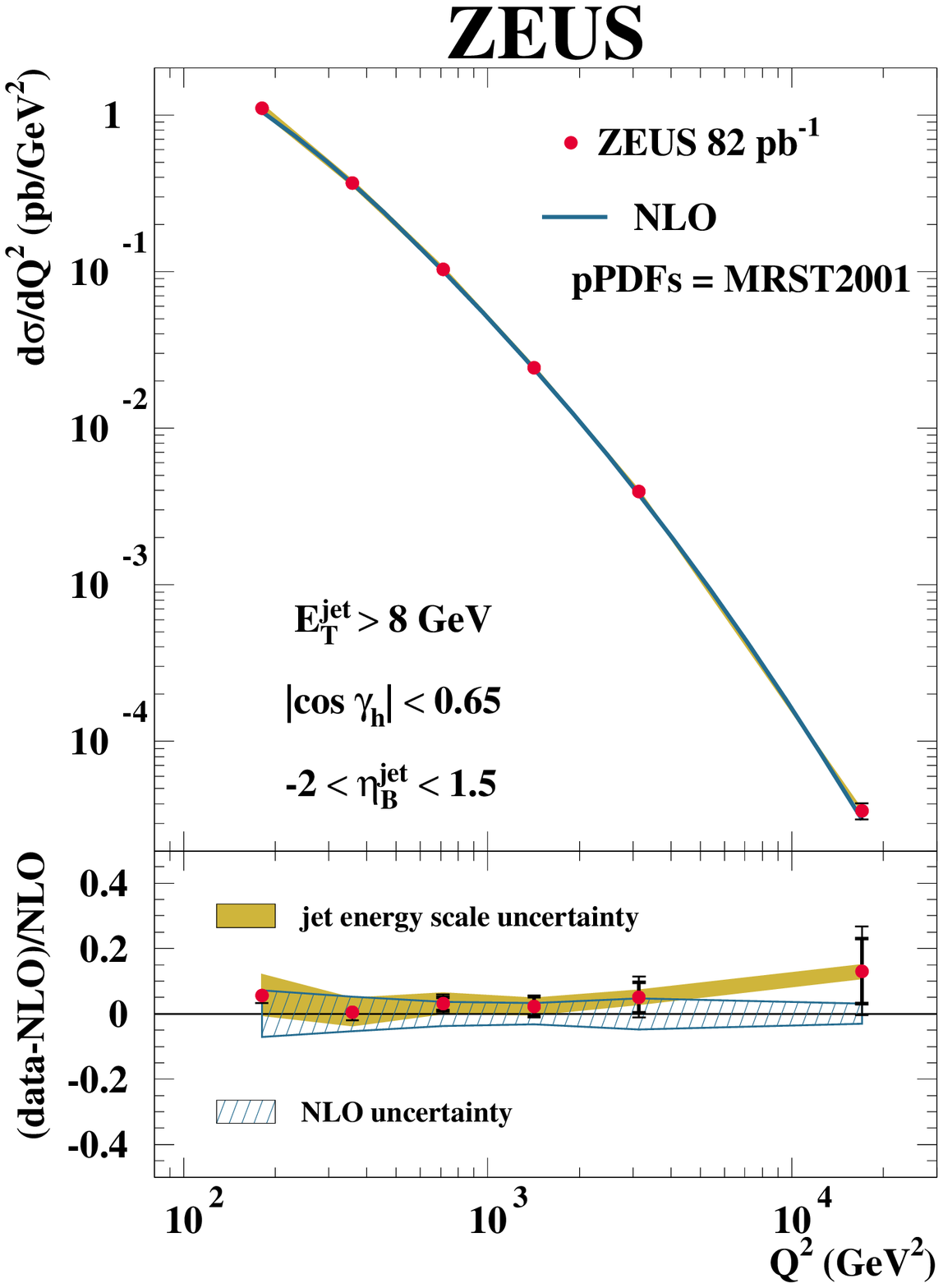,width=8cm}}
\put (8.0,-0.3){\epsfig{figure=\figdir 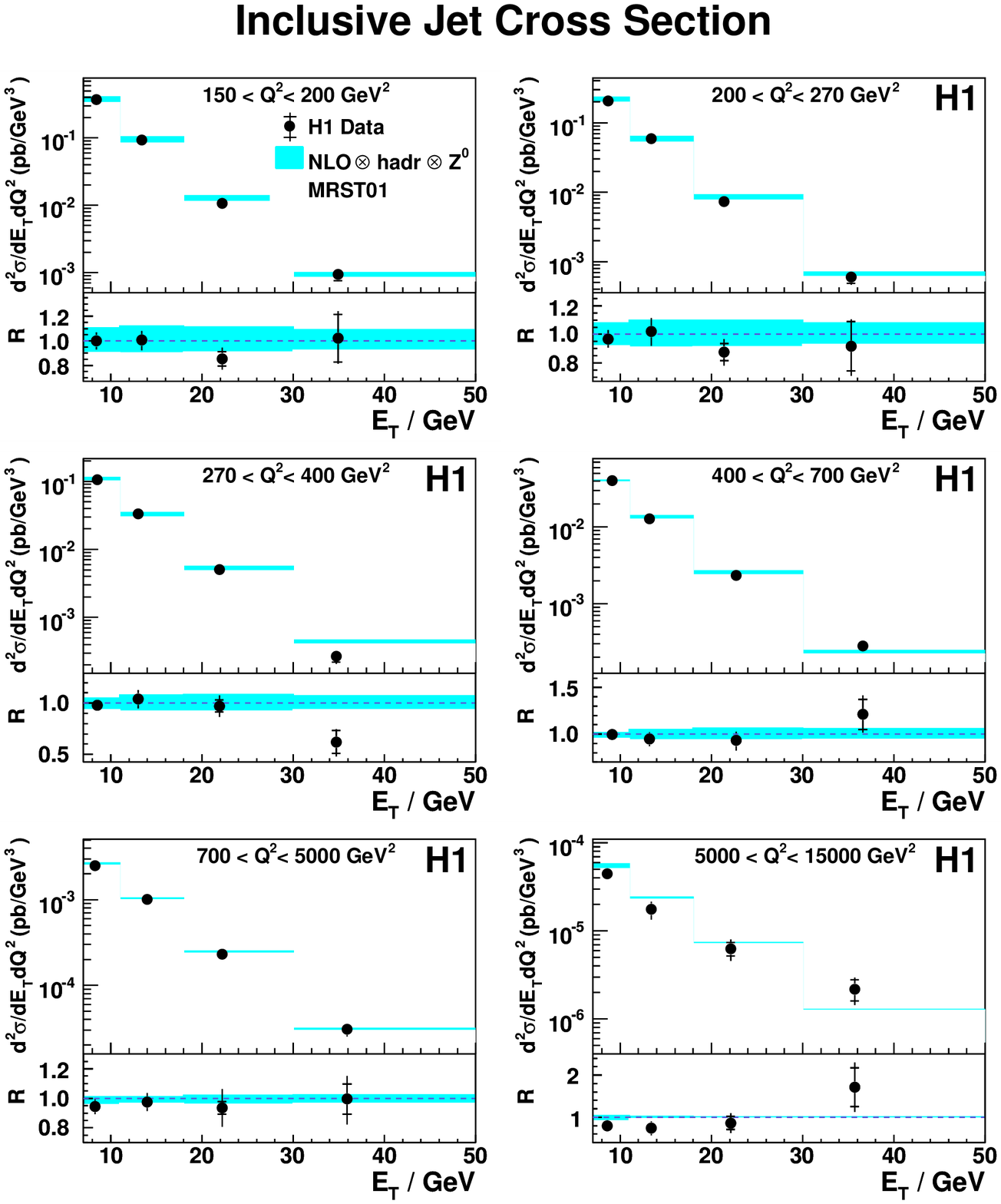,width=6.5cm}}
\put (3.6,-0.3){\bf\small (a)}
\put (11.3,-0.3){\bf\small (b)}
\end{picture}
\caption{\label{fig4}
{
(a) Inclusive-jet single-differential cross-section $d\sigma/d\q2$ as
a function of $\q2$ by ZEUS.
(b) Inclusive-jet double-differential cross-section $d^2\sigma/d\etjb
d\q2$ by H1.
}}
\end{figure}

A combined analysis of H1 and ZEUS data has been performed to obtain
a value of $\asz$. For this, only those measurements which yield the
most precise $\as$ values, namely inclusive-jet cross sections in NC
DIS at high $\q2$, were used. A simultaneous fit to the actual
cross-section measurements, instead of combining $\as$ values as it
was done for the HERA-2004 average, was performed. The data sets
used in the fit are shown in Fig.~\ref{fig4}.

The simultaneous fit was done to 24 H1 data points in the range
$150<\q2<15000$ \g2\ and 6 ZEUS data points in the range
$125<\q2<100000$ \g2. The NLO calculations used were based on the
MRST2001 PDF sets. The renormalisation and factorisation scales were
set to $\etjb$ and $Q$, respectively. The experimental uncertainty on
the combined $\as$ value amounts to 0.0019 and was obtained using the
Hessian method, which fits the sources of systematic uncertainties
such as the energy scale, luminosity, model dependence, etc. The
sources of systematic uncertainty were treated as correlated for
measurements within one experiment, but as uncorrelated between the
two experiments. It was checked that the model dependence, which in
principle could be correlated between experiments, had very little
effect whether it was treated as correlated or uncorrelated. The 
theoretical uncertainty coming from terms beyond NLO was estimated
using the method of Jones et al.~\cite{jones}, and gives the largest
contribution. The other sources of theoretical uncertainty considered
were: PDFs (0.0010), factorisation scale (0.0010) and hadronisation
(0.0004). Therefore, the HERA combined 2007 $\asz$ value is 

\vspace{0.3cm}
\centerline{\fbox{$\asz=0.1198\pm 0.0019\ ({\rm exp.})\pm 0.0026\ ({\rm
      th.})$} (HERA combined 2007).}
\vspace{0.3cm}

This combined value is shown in Fig.~\ref{fig1}b together with the
individual values obtained by both collaborations, the HERA-2004
($0.1186\pm 0.0011 ({\rm exp.})\pm 0.0050 ({\rm th.})$) and
the world-2006 ($0.1189\pm 0.0010$) averages. The measurements are
consistent with each other and with the world average. The HERA-2004
average, which combined many determinations of $\as$, had a very small
experimental uncertainty, but the theoretical uncertainty was 
large. The HERA 2007 combined $\asz$ has a much smaller theoretical
uncertainty, due to the combination of measurements in which the
theoretical uncertainties are well under control, at the expense of a
slight increase in experimental uncertainty. However, the total
uncertainty of the new combined value, $2.7\%$, is smaller than that
of the HERA-2004 average. This value of $\asz$ is very competitive
with the most recent result from LEP~\cite{lep}.

\section*{Acknowledgements}
I would like to thank my colleagues from H1 and ZEUS for their
help in preparing this report. Special thanks to T. Kluge and
J. Terr\oo n.

\end{document}